\documentclass[letterpaper, 10 pt, conference]{ieeeconf} 
\IEEEoverridecommandlockouts           
\usepackage{amsmath}
\usepackage{amsfonts}
\usepackage{graphicx}
\usepackage{fontawesome5}
\usepackage{hyperref}
\usepackage[dvipsnames]{xcolor}
\usepackage{tikz}
\usepackage{booktabs}
\usetikzlibrary{positioning, shapes, fit, calc, backgrounds} 
\tikzset{
    align2above/.style={
        to path={
            (\tikztostart) -- 
            (perpendicular cs: horizontal line through={(\tikztotarget)},
                                 vertical line through={(\tikztostart)})
            \tikztonodes
        }
    },
    align2below/.style={
        to path={
            (perpendicular cs: horizontal line through={(\tikztostart)},
                                 vertical line through={(\tikztotarget)})
            -- (\tikztotarget) \tikztonodes
        }
    }
}

\title{\LARGE \bf
Privacy-Preserving Video Conferencing via Thermal-Generative Images}

\author{Sheng-Yang Chiu$^\ast$, Yu-Ting Huang$^\ast$, Chieh-Ting Lin$^\ast$, Yu-Chee Tseng,\\ Jen-Jee Chen, Meng-Hsuan Tu, Bo-Chen Tung, and YuJou Nieh
\thanks{*Equal contribution.}
\thanks{All the authors are with College of Artificial Intelligence, National Yang Ming Chiao Tung University (NYCU), Taiwan; Yu-Chee Tseng also holds a joint appointment with College of Computer Science, NYCU, and with Miin Wu School of Computing, National Cheng Kung University, Taiwan.}
\thanks{PAIR-LRT-Human dataset is available on \href{https://github.com/acqxi/PAIR-LRT-Human}{GitHub} 
(\url{https://github.com/acqxi/PAIR-LRT-Human})}.
}

\begin{document}

\maketitle
\thispagestyle{empty}
\pagestyle{empty}

\begin{abstract}
Due to the COVID-19 epidemic, video conferencing has evolved as a new paradigm of communication and teamwork. However, private and personal information can be easily leaked through cameras during video conferencing. This includes leakage of a person's appearance as well as the contents in the background. This paper proposes a novel way of using online low-resolution thermal images as conditions to guide the synthesis of RGB images, bringing a promising solution for real-time video conferencing when privacy leakage is a concern. SPADE-SR \cite{Chiu_2023_WACV} (Spatially-Adaptive De-normalization with Self Resampling), a variant of SPADE, is adopted to incorporate the spatial property of a thermal heatmap and the non-thermal property of a normal, privacy-free pre-recorded RGB image provided in a form of latent code. We create a PAIR-LRT-Human (LRT = Low-Resolution Thermal) dataset to validate our claims. The result enables a convenient way of video conferencing where users no longer need to groom themselves and tidy up backgrounds for a short meeting. Additionally, it allows a user to switch to a different appearance and background during a conference.
\end{abstract}


\section{Introduction}
Due to the COVID-19 pandemic, there are growing demands for people to work from home. Online or virtual conferencing has become a major way of communication for teamwork. This requires people to activate webcams and make themselves and their backgrounds visible to everyone else in a video conference room. For this purpose, users must be well-groomed for an extended period of time with proper backgrounds. To facilitate online conferencing, solutions such as Mesh~\cite{Mesh}, Horizon Workrooms~\cite{Horizon_Workrooms}, and Animaze by Facerig have been developed. While attractive, these applications require special equipment such as Microsoft HoloLens 2~\cite{HoloLens} and Meta Quest 2 \cite{Meta_Quest_2}, which may take extensive time to set up. Professional video-based tools dedicated to online conferencing include Zoom, Webex, Meet, and Teams~\cite{singh2020updated}. On the other hand, social tools such as Line, Messenger, and Discord also support conferencing capability.

Most of the aforementioned tools involve the use of built-in cameras, which may easily run into the risk of revealing users' personal information and privacy. Such information may leak out from the users themselves, the backgrounds, or the contents in the background. A significant amount of private information may be compromised either directly from the video contents or indirectly from a side channel by eavesdropping. For example, a device infected with Trojan horses may allow a hacker to capture images from a victim's camera~\cite{pultarova2016webcam}.

\begin{figure} 
      \centering
      \includegraphics[page=6, width=0.47 \textwidth]{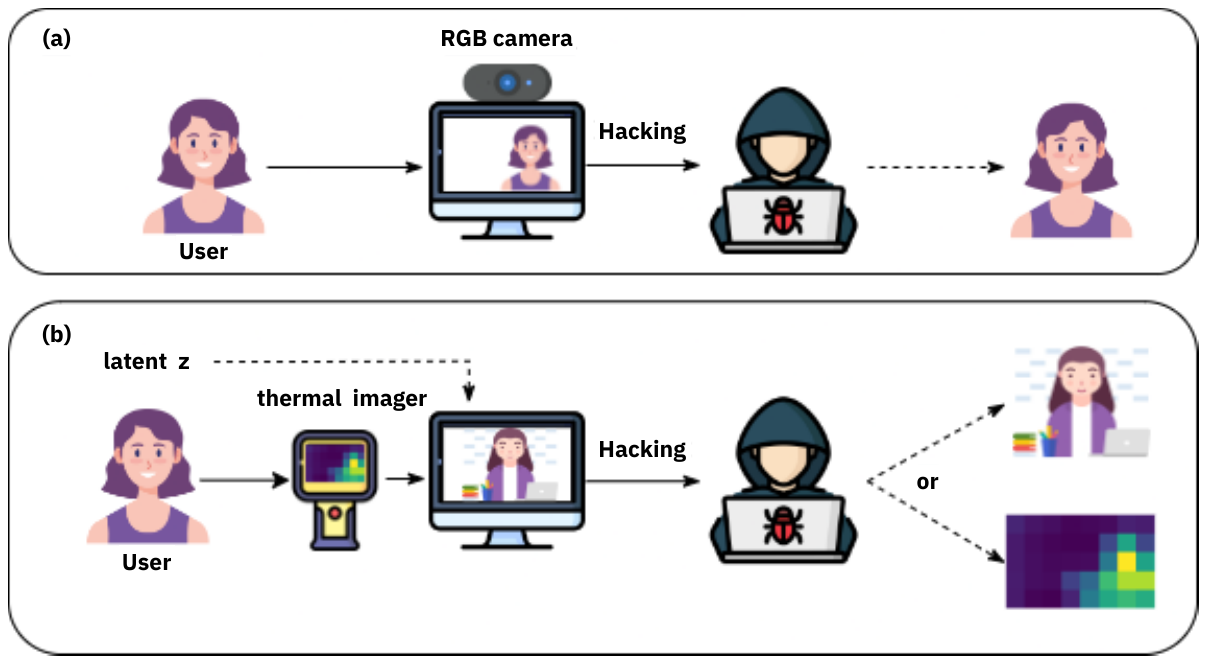}
       \caption{Video conferencing scenarios and potential privacy leakage risks: (a) traditional and (b) thermal-generative.}
       \label{fig:scenarios}
\end{figure}

This work proposes an image generation system for video conferencing. Instead of transmitting real images, we transmit fake real-time images synthesized from (i) a pre-taken, privacy-free image scene provided by the user, which is converted to a form of latent code, and (ii) a sequence of ultra-low-resolution real-time thermal images taken from a thermal sensor. These low-resolution thermal images contain little private information and identifiable objects even if they are compromised.  In addition, it is more convenient because users no longer need to dress up in a tidy home environment for short routine meetings. We compare the traditional and the proposed thermal-generative conferencing scenarios in Fig.~\ref{fig:scenarios}.

In the literature, sensor-based applications have been intensively studied \cite{wang-multiresolution, wu-ban, pan-light-control}. Reference \cite{huang-jin} studied the coverage issue of sensors in a 3D environment. Networking issues of wireless sensors have been discussed in \cite{pan-address}. Recently, several works have applied deep learning techniques for multi-modality sensor data fusion. The work \cite{muka-ieee-sensors} studied person tracking by fusing video data from a drone and wearable sensor data from people on the ground. The work \cite{things-air} addressed cooperative object identification from a drone camera and wireless sensors on the ground. A computer vision-assisted instant alert mechanism was proposed in \cite{tseng-instant-alert}. A new boundary localization scheme through fusing multi-modality sensor data was presented in \cite{liu-boundary}. A novel SensePred framework for wearable sensor-assisted video prediction was proposed in \cite{li-sensepred}. These references motivate us to study fusing sensor and video data for privacy-preserving conferencing.

The rest of this paper is organized as follows. Section II introduces some related works. Section III presents the proposed privacy-preserving video conferencing system. The experiment results are in Section IV. We conclude our work in Section V.

\section{Related Works}
Generative adversarial networks~\cite{goodfellow2020generative} are considered one of the most creative frameworks to produce high-fidelity, high-quality pictures. One of them, DCGANs~\cite{radford2015unsupervised}, which consists of convolutional layers, can generate images from a dataset. Conditional GANs~\cite{mirza2014conditional} extend the field to text-to-image and image-to-image syntheses by controlling some specific labels. AC-GAN~\cite{odena2017conditional} requires only one label and it can generate high-resolution images from noise.

The image-to-image generation has made significant progress due to the works Pix2pix~\cite{isola2017image} and cycleGAN~\cite{zhu2017unpaired}. Pix2pix uses supervised learning with given paired images to generate images and cycleGAN can use a special training cycle to generate an image from different domains by unsupervised learning. Examples include the conversions between different domains, such as painting images to scenery photos~\cite{park2019semantic, huang2021multimodal, brock2018large, wu2019gp}, facial expressions transfer by styleGAN~\cite{karras2020analyzing}, and full-body thermal images to color images~\cite{zhang2018tv}.

With the advance of technologies, using thermal images to produce visible images has become a reality recently~\cite{berg2018generating, damer2019cascaded}. There is considerable progress in thermal infrared image colorization, such as cGAN. Some works try to convert a street view thermal image to a realistic photo~\cite{zhang2018tv} and some transform facial thermal images into realistic human faces~\cite{wang2018thermal, luo2022clawgan, mei2022thermal}. 

On the other hand, in order to generate more realistic images, \cite{park2019semantic, zhu2020sean, sushko2020you, wang2018high} present solutions to generate a synthesized image by a segmentation map. The state-of-the-art model in this issue is Spatial Adaptive Denormalization (SPADE)~\cite{park2019semantic}, which is based on GANs to shift segmentation maps to feature maps. Without relying on segmentation maps, SPADE-SR \cite{Chiu_2023_WACV} is a variant of SPADE designed for transferring low-resolution upper-body thermal images to color images.

Recently, video conferencing has become a popular way of communication. Video conferencing also faces more and more security and privacy risks. The work~\cite{kagan2020zooming} indicates that names, faces, and ages may be leaked through video conferencing. With the progress of face recognition~\cite{schroff2015facenet}, when facing a face recognition attack~\cite{ramachandra2017presentation}, one direct thought is to use GANs to generate appearances of the user with different clothes and expressions. However, the inputs to GANs are still actual images of the user. This motivates us to consider using a latent code of the user combined with a low-resolution thermal image to synthesize images for video conferencing. In this work, we shall adopt SPADE-SR~\cite{Chiu_2023_WACV} to achieve our goal.

\begin{figure}
    \begin{center}
        \scalebox{.8}{\input
            \begin{tikzpicture}[ultra thick, align=center, x=1cm, y=1cm]
                \foreach \x/\y/\sa/\sb/\sc/\t in {
                    1/4/rectangle///{\includegraphics[width=0.12 \textwidth]{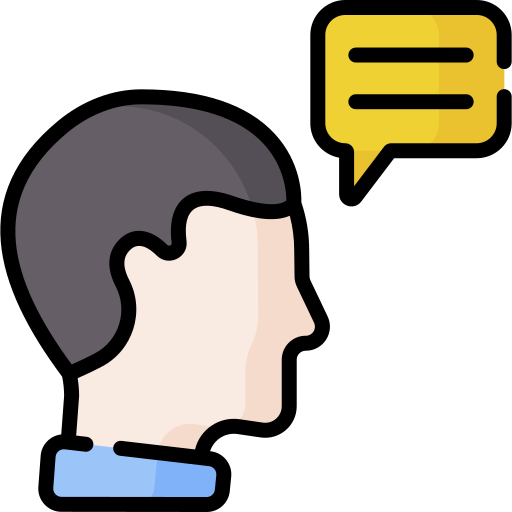}},
                    3/4/ellipse/draw//Thermal\\Sensor,
                    3/2/rectangle/draw//Computer,
                    3/0/ellipse/draw/dotted/RGB\\Camera,
                    6/2/rectangle///,                    
                    8/4/rectangle///outgoing\\synthesized\\images,                    
                    8/2/rectangle///,                    
                    8/0/rectangle///incoming\\images,                    
                }{
                    \node[
                        \sa,
                        \sb,
                        \sc,
                        minimum width=2cm, 
                        minimum height=1.2cm,
                    ] (nG\x\y) at (\x,\y) {\t};
                }
                \draw[->] (nG34.south) -- (nG32.north)
                    node[pos=0.5,left] {$h$};
                \draw[->] (nG30.north) -- (nG32.south)
                    node[pos=0.5,left] {$x$};
                \draw[->, color=Yellow!70, line width=2mm] ($(nG32.east)!0.5!(nG32.north east)$) -| (nG84.south);
                \draw[->, color=RoyalBlue!40, line width=2mm] (nG80.north) |- ($(nG32.east)!0.5!(nG32.south east)$);
                \node[minimum height=1.2cm,draw] at (nG62.center) {Communication\\Interface};
                
            \end{tikzpicture}  
        }
        \caption{Hardware Architecture.}
        \label{fig:hardware}
    \end{center}
\end{figure}
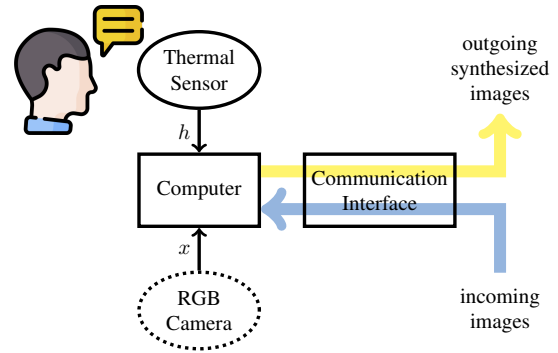

\section{Privacy-Preserving Video Conferencing}

\subsection{Hardware Architecture}
Our hardware architecture is shown in Fig.~\ref{fig:hardware}. It only requires an edge computing client, such as a laptop, that is equipped with a low-resolution thermal imaging sensor. The RGB imaging sensor is optional as it is only required during the training stage to take a photo set representing the user(s). In the online conferencing stage, the user only needs to pick a latent code $z$ from a latent code set $Z$ that represents a pre-taken photo $x$ in the training stage. The edge computing client simply takes $z$ and the current thermal image as inputs and continuously produces synthesized images, which are transmitted to the conferencing partner(s). 

\subsection{Configuring a Video Conference}
\begin{figure} 
      \centering
      \begin{tikzpicture}
        \begin{scope}[shift={(\linewidth,0)},local bounding box=A]
            \node[anchor=north west, inner sep=0pt] (image) at (0,0) {
                \includegraphics[page=8, width=0.47 \textwidth]{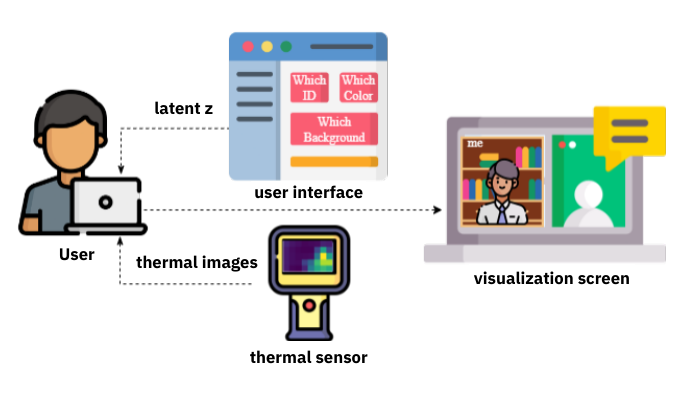}};
        \end{scope}
    \end{tikzpicture}
    \caption{Setting up a privacy-preserving video conference.}
    \label{fig:appl}
\end{figure}

The process to configure a video conference is shown in Fig.~\ref{fig:appl}. From the pre-taken photo set, the user can pick the appearance and background to his preference. The selection is mapped to a latent code $z \in Z$. Then the thermal sensor continuously streams thermal images to the laptop, which will convert them to fake images on the conferencing screen. 

\begin{figure} 
    \centering
    \includegraphics[page=4, width=0.47 \textwidth]{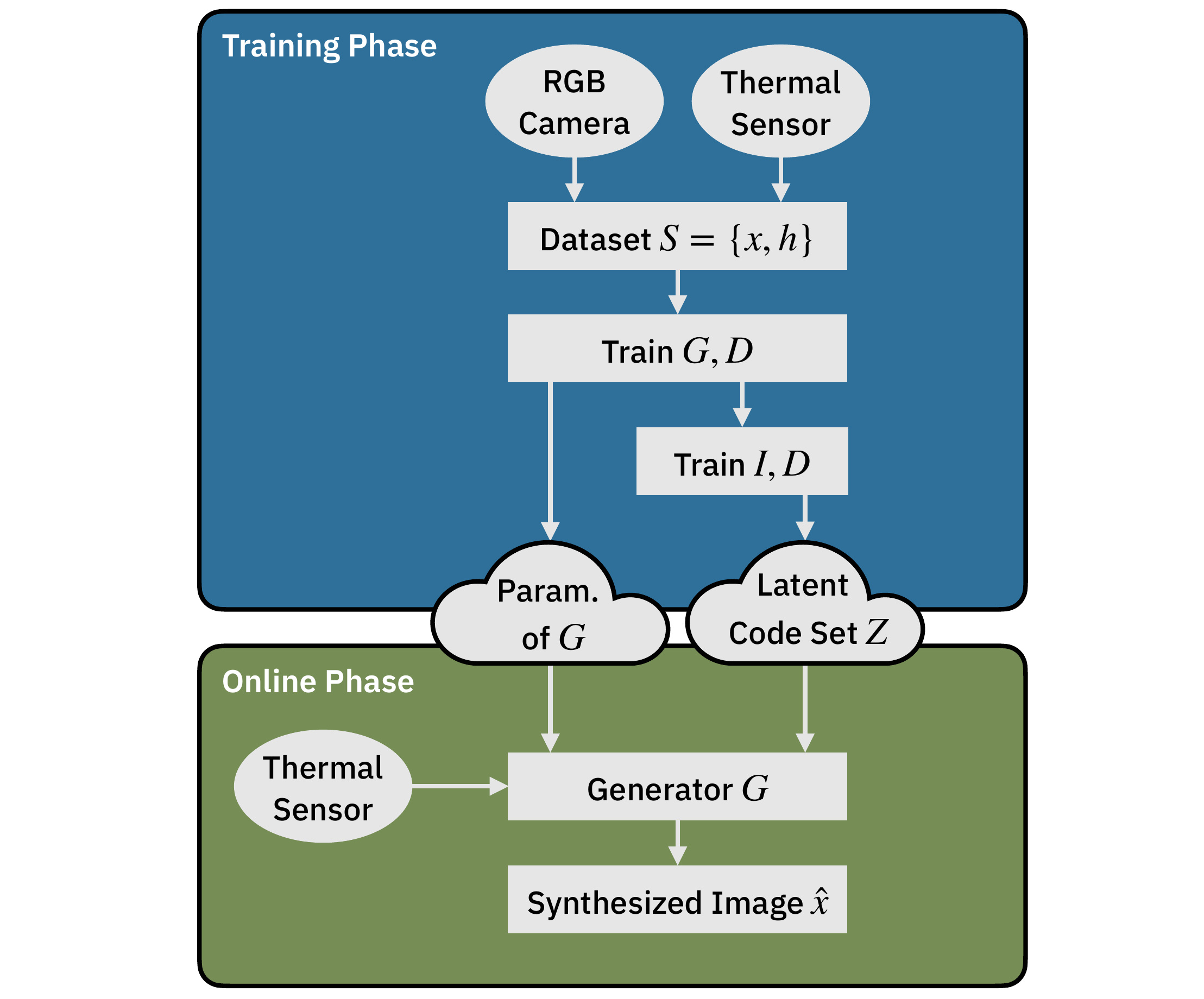}
    \caption{Training and online phases.}
    \label{fig:training}
\end{figure}

The deep learning model includes a training phase and an online phase, as shown in Fig.~\ref{fig:training}. In the training phase, the first step is to collect a dataset $S$ that consists of a number of pairs $(x, h)$, where $x$ is an RGB image and $h$ is a thermal image that are taken simultaneously. Then we train our thermal-GAN model (refer to the next subsection), which includes training $G$ and $D$ of GANs and training $D$ and $I$ of Inversion. The outputs include (i) the parameters of $G$ and (ii) a latent code set $Z = \{ \tilde z| \tilde z = I(x), (x, \cdot) \in S \}$. That is, $Z$ is obtained by applying $I$ on each $x$ in the dataset $S$.

\subsection{Thermal-GAN}

Our aim is to produce a high-quality RGB image $\hat{x}$ using a low-resolution heatmap $h$ acquired from a thermal sensor. The synthesized image $\hat{x}$ should be as close to the original image $x$ as possible such that $\left\{x, h\right\} \in S$. To accomplish this, we will use GANs, specifically the SPADE-SR (SPADE with Self-Resampling) model \cite{Chiu_2023_WACV}, to build our video conferencing system. This model takes a thermal heatmap $h$ and a latent code $z\sim P_{noise}(z)$ as inputs and produces an image $\hat{x}=f(z, h)$ by following the conditional GAN paradigm, or more precisely, the AC-GAN, where $h$ serves as a condition. SPADE-SR follows AC-GAN since the thermal sensor data has a very low resolution, which can safeguard users' privacy. Moreover, sensor data is often quite noisy. The design not only recognizes the relationship between $h$ and $x$, but also encourages high mutual information between the generated image $\hat{x}$ and $h$ \cite{chen2016infogan}.

\begin{figure} 
      \centering
      \includegraphics[page=1, width=0.47 \textwidth]{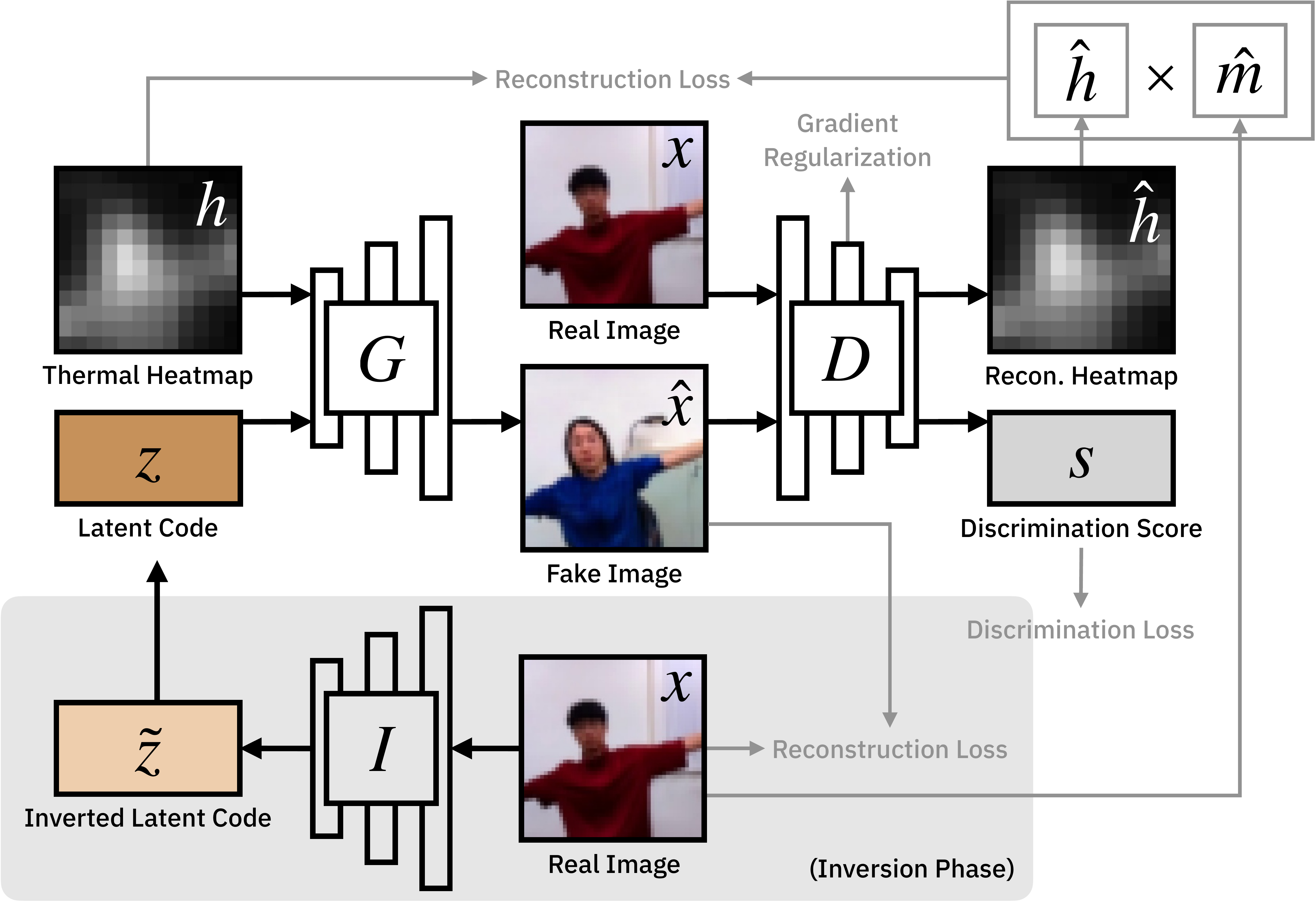}
       \caption{The SPADE-SR model for privacy-preserving image synthesis.}
       \label{fig:model}
\end{figure}

Fig.~\ref{fig:model} shows the architecture of SPADE-SR \cite{Chiu_2023_WACV}. The dataset is denoted as \(S=\{d_1,d_2,\dots,d_K\}\), where each \(d_i=(x_i,h_i),i=1\dots K\), consists of an RGB image \(x_i \in {R}^{H_x\times W_x\times 3}\) and a thermal heatmap \(h_i \in {R}^{H_h\times W_h\times 1}\). $G$ generates an RGB image \(\hat{x}=G(z,h) \in {R}^{H_x \times W_x \times 3}\), where \(h\) is a heatmap and \(z\sim \mathcal{N}(0,1),z \in {R}^{256}\), is a noise variable representing non-thermal attributes independent of \(h\).
The discriminator \(D\) takes an image and outputs a reconstructed heatmap \(\hat{h} \in {R}^{H_h\times W_h\times 1}\) and a realness score \(s \in {R}^1\). 
$I$ is an inversion encoder that converts a source image \(x\) to its matching non-thermal code $\tilde{z}$. The code $\tilde{z}=I(x)$ is then used in conjunction with other heatmaps to synthesize more RGB images.

\begin{figure} 
      \centering
      \includegraphics[page=2, width=0.47 \textwidth]{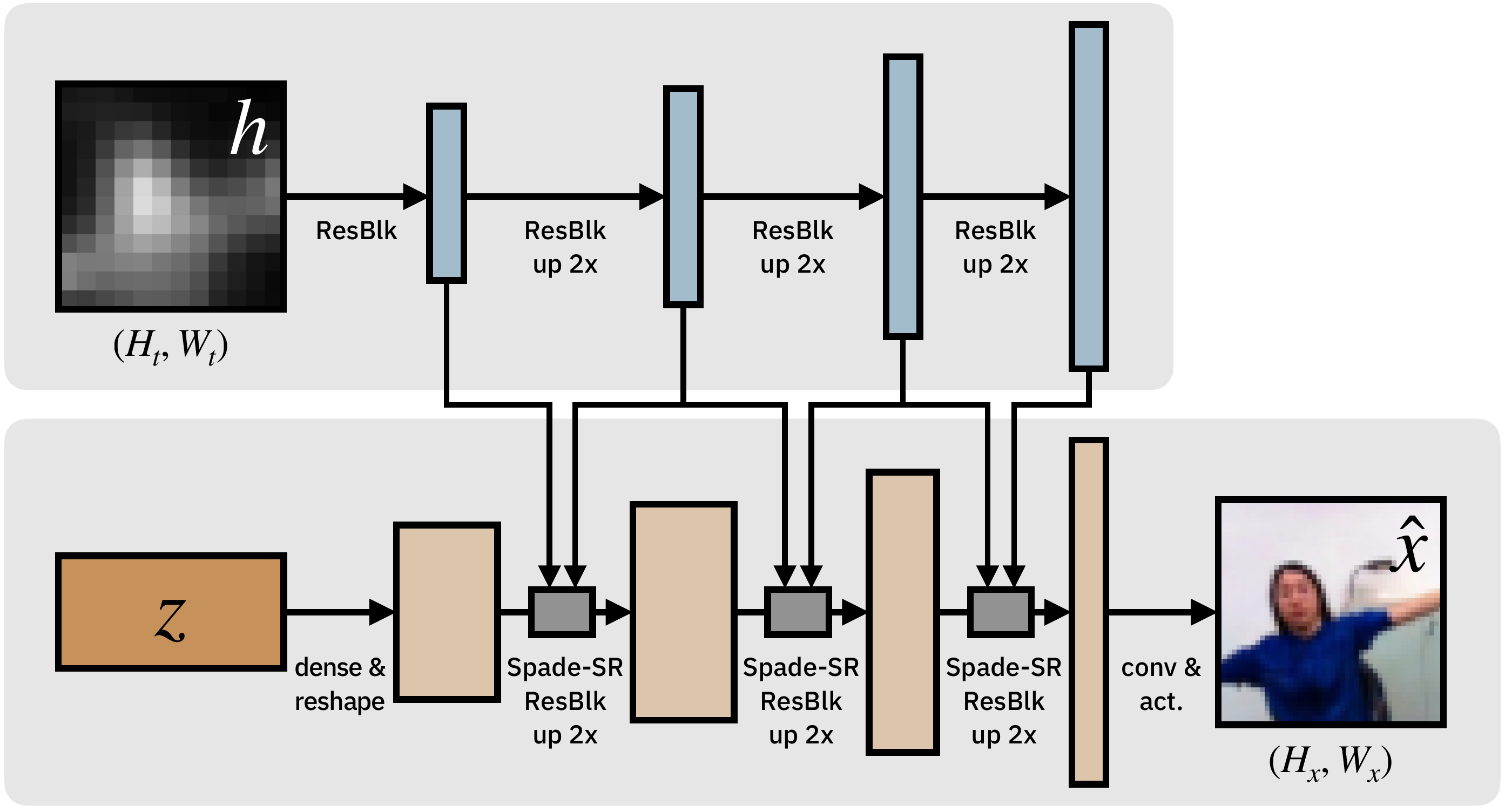}
       \caption{The generator $G$ and the resampling process to produce $\hat{x}$.}
       \label{fig:GENERATOR_STRUCTURE}
\end{figure}

Fig.~\ref{fig:GENERATOR_STRUCTURE} shows the architecture of $G$. It is made up of two network branches. The main branch is composed of SPADE-SR ResBlock layers \cite{Chiu_2023_WACV} that successively convert the noise vector \(z\) into an RGB image \(\hat{x}\). To process the thermal heatmap \(h\), the semantic branch continuously resamples itself to a larger feature map through ResBlocks. There are three phases of upsampling (three SPADE-SR ResBlocks), so the generated image \((H_x,W_x)\) is eight times larger than $(H_t,W_t)$.

\subsection{Loss Functions}
We propose three loss functions to facilitate the training of our model. The first one pertains to the reconstruction loss between the original input feature $h$ and its predicted counterpart $\hat{h}$. The second loss function relates to the discrimination loss of the discriminator $D$, while the third one corresponds to the reconstruction loss between the original RGB image $x$ and its predicted version $\hat{x}$. The latter two loss functions are based on the definitions presented in SPADE-SR \cite{Chiu_2023_WACV}.

For the first reconstruction loss, a mask ${m}$ is obtained from the RGB ground truth image $x$ after performing person segmentation. The mask $m$ has a size of ($H_x, W_x$). We resize it to the size of the thermal image ($H_t, W_t$), denoted as $\hat{m} = [\hat{m}_{i, j}]$. We then apply $\hat{m}$ on $h$ and $\hat{h}$ and compare the results using the hinge L1 loss, which ensures that $\hat{h}$ is similar to $h$ while limiting the deviation no more than a small $\epsilon$. This can be expressed as:
\begin{equation} 
    \label{eq:7}
    \begin{aligned}
        L_{\mathrm{rec}} = \sum_{i,j} \mathbb{I}(\hat{m}_{i,j} = 1) \cdot \left[ \max(0, |h_{i,j} - \hat{h}_{i,j}| - \epsilon)\right]
        .
    \end{aligned}
\end{equation}
Here, $h_{i,j}$ and $\hat{h}_{i,j}$ represent the pixel values at position $(i,j)$ in $h$ and $\hat{h}$, respectively. The function $\mathbb{I}(\cdot)$ is an indicator function that is equal to 1 when the condition in parentheses is true, and 0 otherwise. The mask enforces the model to focus only on the part with human thermal.

\section{Experiment Results}
\subsection{Dataset}
\subsubsection{Data Collection Process}
\begin{figure} 
    \centering
    \includegraphics[page=5, width=0.47 \textwidth]{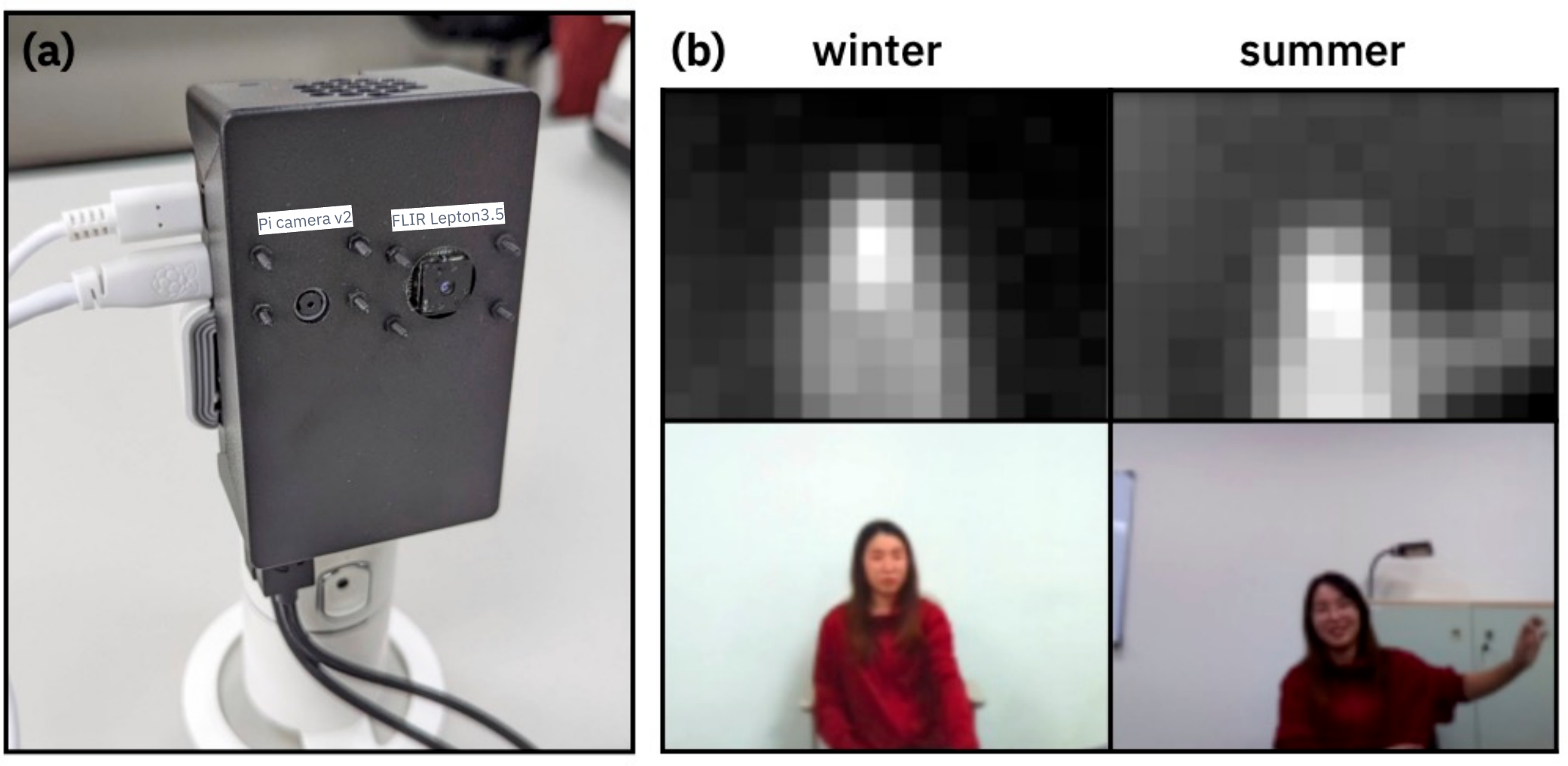}
    \caption{(a) Hand-made data collection device based on Raspberry Pi 4B. (b) Recording environment and two data samples. There was a bigger temperature difference between the subject and the background during winter.}
    \label{fig:sensor_diff}
\end{figure}

We have created a PAIR-LRT-Human (PAIR Lab Low-Resolution Thermal-Human images) dataset. The collection device is a hand-crafted Raspberry Pi with a FLIR Lepton3.5 thermal camera and a Pi camera v2, as shown in Fig.~\ref{fig:sensor_diff}(a). Before collecting data, fisheye calibration on the camera is required, followed by pairing the camera with the thermal sensor to ensure that there is as much overlapping between their views as possible. Because the thermal sensor has a lower rate of \(8.7\) FPS, we regulate the sampling rate of the pi camera. That is, whenever a thermal image is recorded, we read out an RGB image.
 
The experiment was kept with as few irrelevant heat sources as possible. There was only one person moving into the area. The individual must face the device while wearing light clothes and doing a wide range of actions. The room temperature was under control; however, as can be seen from the figure, the background wall temperatures were still different in summer and winter. This dataset includes two persons, two scenarios, and three clothing colors, for a total of \(12\) different settings. Each environment has up to 11K thermal-RGB image pairs, for a total of approximately 33K pairs.

\subsubsection{Data Preprocessing}
\begin{figure} 
      \centering
      \includegraphics[page=9, width=0.47 \textwidth]{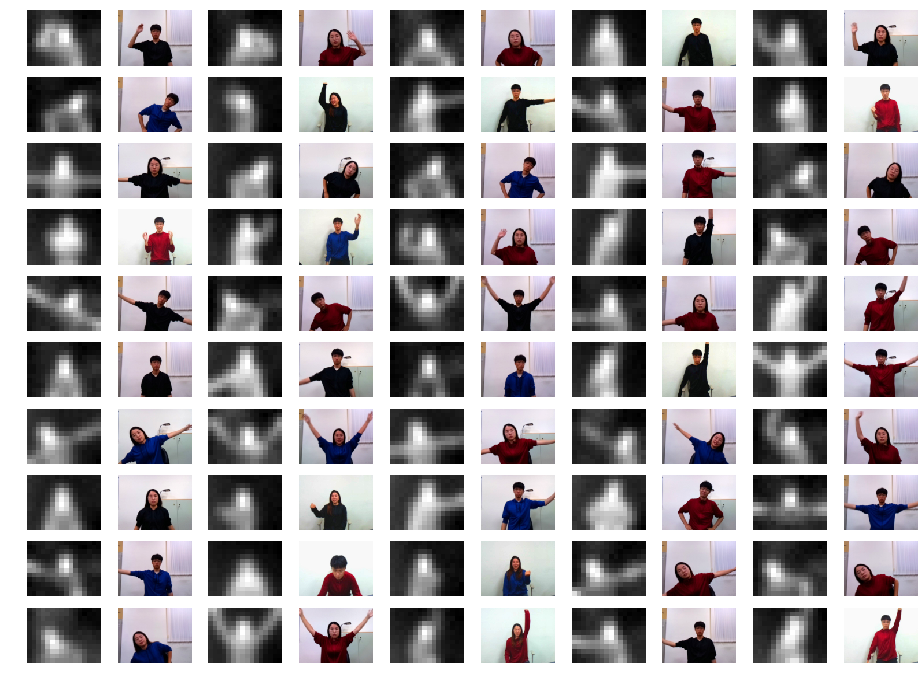}
       \caption{Random samples of thermal-RGB pairs from the PAIR-LRT-Human dataset after data pre-processing.}
       \label{fig:random}
\end{figure}

We aim to align the two types of images. For each RGB image, we down-sample it to (H=96,W=128,C=3) by pixel averaging and then normalize each value to $[-1, 1]$. For each thermal image, we blur it using a Gaussian filter with kernel size \(\alpha\)~\cite{deriche1993recursively}, add random Gaussian noise \cite{goodman1963statistical}, down-sample it via pixel averaging \cite{garcia2010pixel} to (H=12, W=16, C=1), and normalize each value to $[-1, 1]$. The down-sampling step is to ensure that the thermal images have a very low resolution that contains almost no meaningful information for human eyes. Our goal is to synthesize RGB images 8 times larger than thermal images. Fig.~\ref{fig:random} contains some random samples in PAIR-LRT-Human with various human poses and positions under various clothing and lighting situations. 

A comparison between PAIR-LRT-Human and LRT-Human created by \cite{Chiu_2023_WACV} is shown in Table~\ref{table:DATASET_SPEC}.

\begin{table}
    \setlength{\tabcolsep}{2pt}
    \caption{Comparison of PAIR-LRT-Human and LRT-Human Datasets.}
    \label{table:DATASET_SPEC}
    \centering
    \scalebox{1.2}{
        \input
        \begin{tabular}{{lll}}
            \toprule
            \midrule
            & PAIR-LRT-Human & LRT-Human \\
            \midrule
            thermal sensor & FLIR Lepton3.5 & Pana. AMG8833 \\
            heatmap resolution & \(16\times 12\) & \(8\times 5\) \\
            RGB camera & Pi camera v2 & iPhone XS \\
            RGB resolution & \(128\times 96\) & \(64\times 40\) \\
            field of view & $71^\circ \times 57^\circ$ & $60^\circ \times 37.5^\circ$ \\
            No of data pairs & (33, 228) & (21, 959) \\
            \midrule
            human occupant & 1 & none or 1 \\
            standing position & random & random \\
            pose & upper-body & upper-body \\
            clothing colors & 3 & 5 \\
            env conditions & 3 & 2 \\
            No of persons & 2 & 1 \\
            \midrule
            \bottomrule
        \end{tabular}
    }
\end{table}

\subsection{Validations}
\subsubsection{Model Training and Testing}
We adjust dimensions $(H_x,W_x)$ of $x$ and $(H_h,W_h)$ of $h$ to match the dataset specification. Our goal is to achieve $H_x = 8 \times H_h$ and $W_x = 8 \times W_h$. To train the GAN, the Adam~\cite{kingma2014adam} optimizer is used with \(\beta_1=0\), \(\beta_2=0.999\), and learning rates \(=2 \times 10^{-4}\) and \(5 \times 10^{-5}\) for \(D\) and \(G\), respectively. During the inversion phase, we utilize Adam with \(\beta_1=0\), \(\beta_2=0.999\), and learning rates \(=2 \times 10^{-4}\) and \(5 \times 10^{-5}\) for the \(D\) and \(I\), respectively. The batch size is set to \(128\), and the update ratios for \(D\) to \(G\) (GAN) and for \(D\) to \(I\) (inversion) are both 1. We set 400 epochs to train the GAN phase, and 50 epochs to train the inversion phase. We apply exponential moving averages on model parameters for evaluation using decay rates of \(0.999\) and \(0.99\) for the GAN and inversion phases, respectively. The total training time for GAN and inversion phases takes about 26.5 hours on four Nvidia V100 GPUs. The image generation rate reaches 11.7 fps with one single input and up to 389 fps with multiple inputs. 

\subsubsection{Generation Quality}
We measure the RGB image synthesis quality by FID (Fréchet Inception Distance) \cite{heusel2017gans}. FID calculates the distance between two feature maps extracted by the Inception network from two input images. It helps measure the similarity of two images when they are describing the same person. It is important to note that the feature maps used are retrieved from a specific layer of the Inception network, which are in the form of a 1D vector of length 2048. The FID score is defined as follows:
\begin{equation}
FID=\left\lVert\mu_{1}-\mu_{2}\right\rVert_{2}^{2}+Tr\left(\Sigma_{1}+\Sigma_{2}-2\left(\Sigma_{1}\Sigma_{2}\right)^{\frac{1}{2}}\right)
,
\end{equation}
where $\mu_{1}$ and $\mu_{2}$ are the mean vectors, and $\Sigma_{1}$ and $\Sigma_{2}$ are the covariance matrices of the feature maps of the original and the generated reference sets of images, respectively. $Tr$ denotes the trace operator and $\left\lVert\cdot\right\rVert_{2}$ denotes the Euclidean norm. Since FID calculates the distance between two one-dimensional vectors, a smaller distance or a lower FID score indicates that the two vectors are closer, which means that the corresponding images are more similar. As the result, our method achieves an FID score of 9.6.

\begin{figure} 
      \centering
      \includegraphics[width=0.45 \textwidth]{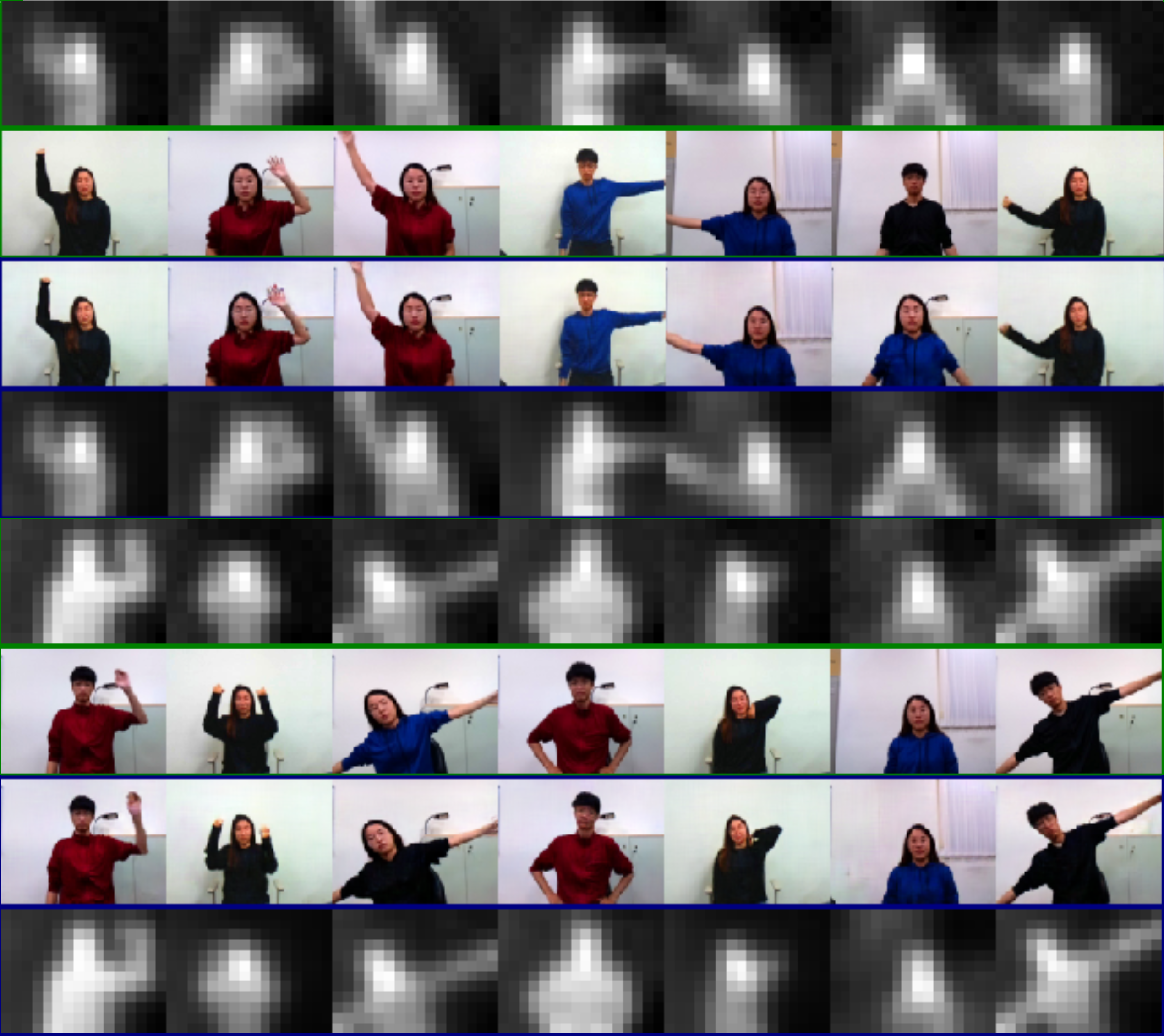}
       \caption{Some samples of ground truth and generated images. Green rows are random samples from PAIR-LRT-Human, and blue rows are synthesized RGB and thermal images.}
       \label{fig:REALandFAKE}
\end{figure}

To visualize the synthesis quality, we show some examples in Fig. \ref{fig:REALandFAKE}. We randomly select a portion of the dataset as the test set and use the heatmap of a test data pair and the corresponding person's latent code as input. We compare the generated RGB image and heatmap with the ground truth. The results demonstrate that our method is capable of producing results that are visually similar to the ground truth.

\subsubsection{Avoiding Entanglement}
\begin{figure*} 
      \centering
      \includegraphics[page=11, width=0.99 \textwidth]{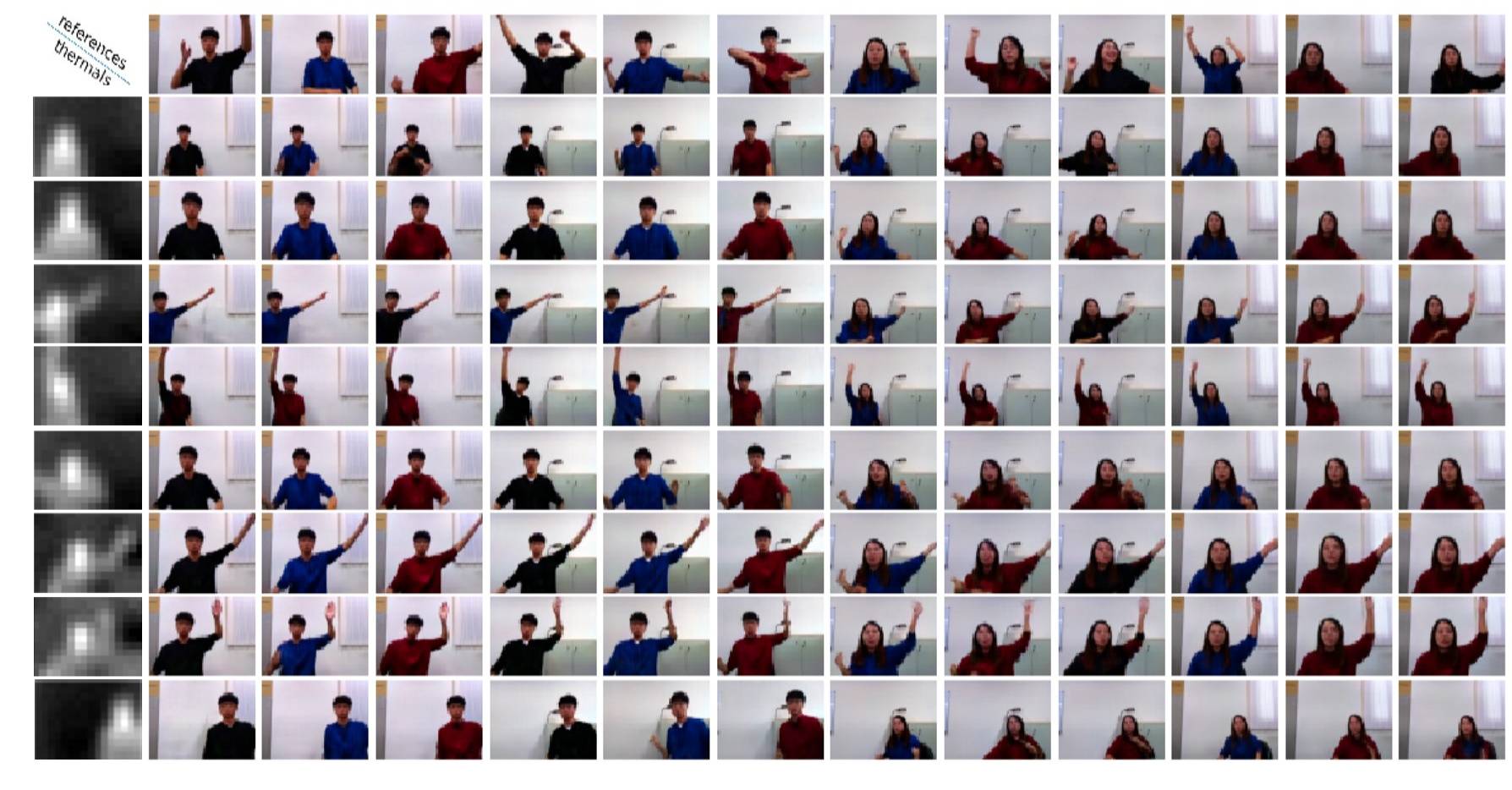}
       \caption{Disentanglement evaluation. The top row shows the source images and the leftmost column shows the heatmaps. Each cell represents a generated image given the corresponding non-thermal code and heatmap.}
       \label{fig:fid}
\end{figure*}

We synthesize RGB images from thermal and RGB image features. It is expected that these features are not entangled, which implies that each feature is only relevant to the real features that have been ascribed to it. To evaluate the disentanglement performance, we randomly pick \(k\) RGB images from the dataset and use the inversion encoder to get the features of RGB images; then we randomly sample \(n\) thermal images and use them to produce \(k\times n\) images. We place these produced images in a \(k\times n\) grid to see how one attribute changes graphically under the influence of the other attribute. As shown in Fig.~\ref{fig:fid}, these two attributes have a clear and independent relationship - the heatmap controls the pose and position of the person and the latent code affects the clothing, background, and gender information.

\subsubsection{Privacy Protection}
With respect to the trade-offs between thermal image resolution and privacy protection, in Table \ref{table:PRIVACY_PROTECTION} we evaluate privacy protection by using a fine-tuned YOLO v7 model \cite{wang2022yolov7} to detect people in a thermal image. Intuitively, if a person in a thermal image can be detected by the model, there is a higher risk that some details in the thermal image may reveal the person's private information. Therefore, the higher accuracy in detecting people in thermal images means a lower degree of privacy protection. Conversely, a lower accuracy in detecting people indicates a higher degree of privacy protection. Our experiment results show that when using the fine-tuned YOLO v7 model on thermal images of a resolution of \(160\times 120\), we achieve \(80.7\%\) accuracy. However, for thermal images with resolutions of \(16\times 12\) and \(8\times 5\), the accuracy is \(0\%\). This supports our claim that reducing the thermal image resolution can increase privacy protection.

\begin{table}
  \setlength{\tabcolsep}{12pt}
  \caption{Privacy protection degree comparison.}
  \label{table:PRIVACY_PROTECTION}
  \centering
  \begin{tabular}{{lcc}}
    \toprule
    \midrule
    Thermal Image & YOLO v7 & Privacy Protection \\ Resolution & Accuracy & Degree
    \\
    \midrule
    160x120 & 80.7\% & Low \\
    16x12 & 0\% & High \\
    8x5 & 0\% & High \\
    \midrule
    \bottomrule
  \end{tabular}
\end{table}

\subsubsection{Demonstration}
A live video demonstration is provided on the Internet\footnote{A demo video is available in \href{https://youtu.be/angJdjjt3Fs}{YouTube} (\url{https://youtu.be/angJdjjt3Fs})}.
Fig.~\ref{fig:failure} reports some failure cases.

\subsubsection{Discussion and Limitations}Below, we discuss some limitations of our current work.

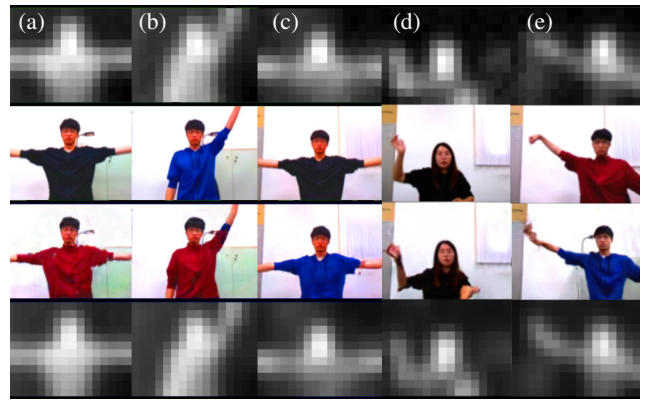
\begin{figure} 
    \centering
    \begin{tikzpicture}
        \begin{scope}[shift={(0.0\linewidth,0)},local bounding box=A]
            \node[anchor=north west, inner sep=0pt] (image) at (0,0) {\includegraphics[width=0.47 \textwidth]{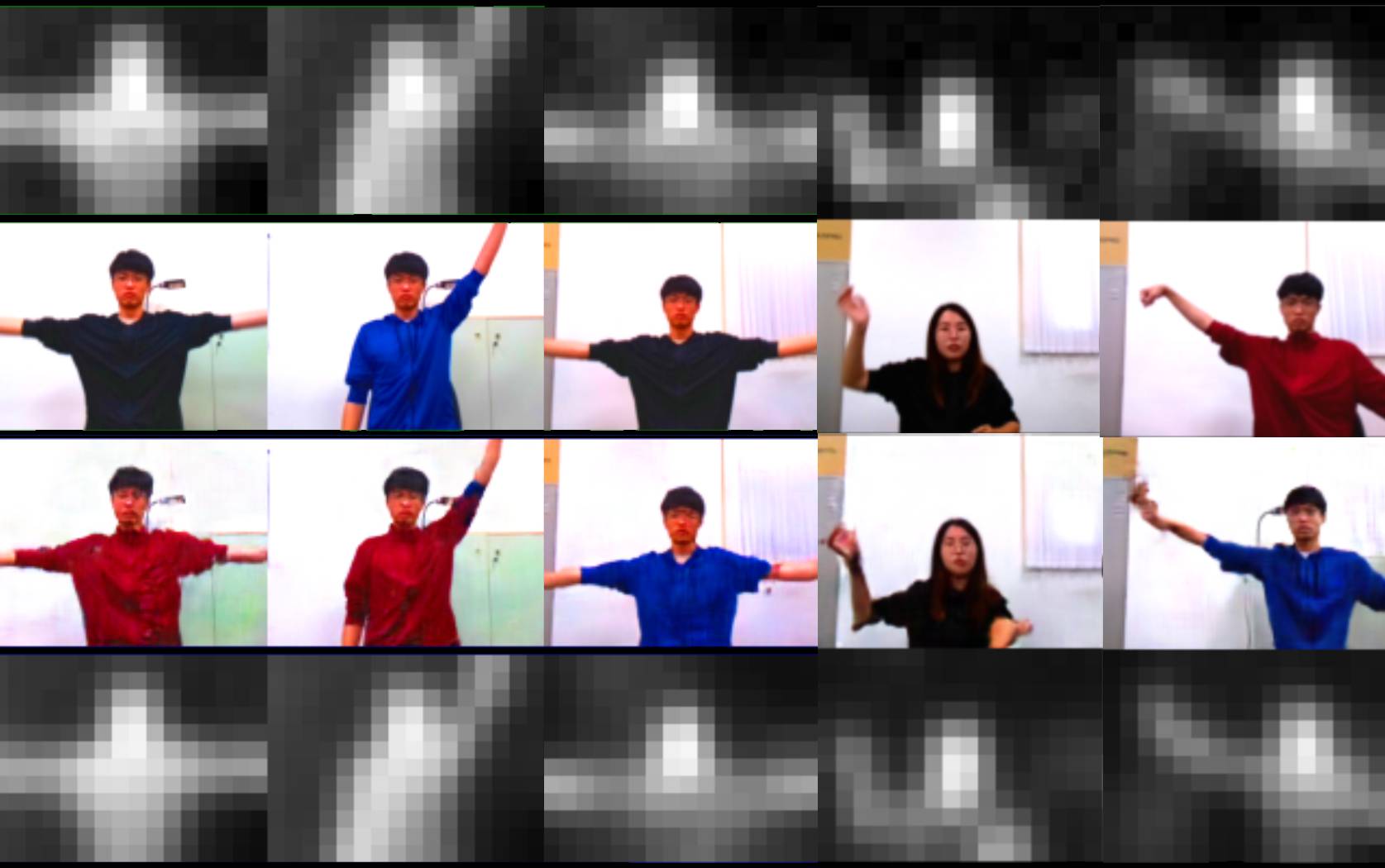}};
            \node[anchor=north west, inner sep=0pt, text=white] at ([xshift=3pt,yshift=-3pt]image.north west) {\small (a)};
            \node[anchor=north west, inner sep=0pt, text=white] at ([xshift=3pt+0.09 \textwidth,yshift=-3pt]image.north west) {\small (b)};
            \node[anchor=north west, inner sep=0pt, text=white] at ([xshift=3pt+0.19 \textwidth,yshift=-3pt]image.north west) {\small (c)};
            \node[anchor=north west, inner sep=0pt, text=white] at ([xshift=3pt+0.28 \textwidth,yshift=-3pt]image.north west) {\small (d)};
            \node[anchor=north west, inner sep=0pt, text=white] at ([xshift=3pt+0.38 \textwidth,yshift=-3pt]image.north west) {\small (e)};
        \end{scope}
    \end{tikzpicture}
    \caption{Failure cases. (a)(b)(c) are speculated to be caused by a small number of images that fail to generate correct colors due to the inability of reconstruction loss in the inverse stage to accurately identify colors. (d) may be due to the fact that the model can only verify its own results by regenerating low-resolution thermal images, so it is not too important whether the arm shape is correct. (e) is a complex pose that may cause motion errors due to insufficient learning.}
    \label{fig:failure}
\end{figure}

\begin{itemize}
    \item 
    Since we only use very low-resolution thermal images for image generation, it is still quite difficult to synthesize images with sufficient facial details. However, if higher-resolution thermal sensors are used, more personal information may be disclosed. This is a dilemma. Future work may consider using higher resolutions for facial expressions synthesis and lower resolutions for upper body synthesis.
    \item
    While our system performs well, there are still some failure cases in our current results, which are shown in Fig.\ref{fig:failure}. We shall undergo more data collection and data augmentation to improve our dataset. For example, data collection in different environments with various backgrounds and external thermal sources can be conducted.
    \item
    Currently, to run our video conferencing system, one has to pick a latent code $z$ in the existing latent code set $Z$. These latent codes are pre-recorded. That is, one is not able to use an image not pre-trained by our system. To do so, the model needs to be retrained.
    \item
    Our system has been validated with a small set $Z$. That is, it is more suitable for use by a single person or a small working group. How to scale up its size deserves further study.
\end{itemize} 

\section{CONCLUSIONS}
In this paper, we have addressed the privacy protection problem of video conferencing. To tackle this challenge, we propose an image generation system using thermal sensors based on the state-of-the-art method SPADE-SR. We also propose to improve the loss function of SPADE-SR by using a detected human mask on the heatmap to separate external thermal sources from human thermal sources. What this system actually uploads to the Internet are synthetic images, so even if the system is hacked, our system is more secure in terms of privacy protection. The privacy protection capability is validated via a human detection model that tries to identify people from a thermal image. We evaluate our model on the newly proposed PAIR-LRT-Human dataset and it demonstrates excellent image generation quality with an FID score of 9.6  and disentanglement properties. 

In terms of synthesis speed, our model reaches 11.7 fps with one single input, and up to 389 fps with multiple inputs.

\addtolength{\textheight}{-2cm}  

\section*{ACKNOWLEDGMENT}
This work was jointly sponsored by the National Science and Technology Council (NSTC), Taiwan, under grant number NSTC 111-2218-E-A49-012-MBK, and ``Center for Open Intelligent Connectivity’’ of ``Higher Education Sprout Project’’ of MOE, Taiwan.


\bibliographystyle{IEEEtranBST/IEEEtran}
\bibliography{IEEEtranBST/ICRA}

\end{document}